\documentclass{article}
\usepackage{spconf,amsmath,graphicx,amsfonts,cite,tabularx,multirow,xcolor,url,breakurl}
\usepackage[breaklinks]{hyperref}

\newcolumntype{C}{>{\centering\arraybackslash}m{5ex}}
\newcolumntype{D}{>{\arraybackslash}m{42ex}}
\newcolumntype{E}{>{\centering\arraybackslash}m{6ex}}
\newcolumntype{F}{>{\centering\arraybackslash}m{5ex}}

\title{FROM SENONES TO CHENONES: TIED CONTEXT-DEPENDENT GRAPHEMES FOR HYBRID SPEECH RECOGNITION \thanks{\textit{* To appear at ASRU 2019}}}
\name{Duc Le, Xiaohui Zhang, Weiyi Zheng, Christian F{\"u}gen, Geoffrey Zweig, Michael L. Seltzer}
\address{Facebook AI\\
	{\small \texttt{\{duchoangle,xiaohuizhang,wyz,fuegen,gzweig,mikeseltzer\}@fb.com}}}
\begin{document}
	%
	\maketitle
	\begin{abstract}
		There is an implicit assumption that traditional hybrid approaches for automatic speech recognition (ASR) cannot directly model graphemes and need to rely on phonetic lexicons to get competitive performance, especially on English which has poor grapheme-phoneme correspondence. In this work, we show for the first time that, on English, hybrid ASR systems can in fact model graphemes effectively by leveraging tied context-dependent graphemes, i.e., \textit{chenones}. Our chenone-based systems significantly outperform equivalent senone baselines by 4.5\% to 11.1\% relative on three different English datasets. Our results on Librispeech are state-of-the-art compared to other hybrid approaches and competitive with previously published end-to-end numbers. Further analysis shows that chenones can better utilize powerful acoustic models and large training data, and require context- and position-dependent modeling to work well. Chenone-based systems also outperform senone baselines on proper noun and rare word recognition, an area where the latter is traditionally thought to have an advantage. Our work provides an alternative for end-to-end ASR and establishes that hybrid systems can be improved by dropping the reliance on phonetic knowledge.
	\end{abstract}
	\section{INTRODUCTION}
	\label{sec:intro}
	
	In the past decade, neural network acoustic models have become a staple in automatic speech recognition (ASR). This movement began with the hybrid approach where Hidden Markov Model (HMM) models the temporal property of speech and Deep Neural Network (DNN) replaces Gaussian Mixture Model (GMM) to estimate emission probabilities of HMM states \cite{Dahl2011,DBLP:journals/taslp/MohamedDH12,Dahl:2012:CPD:2335874.2336015,Hinton2012}. DNN is subsequently replaced with variants of Recurrent Neural Network (RNN) and Convolutional Neural Network (CNN) \cite{GravesMH13,DBLP:conf/interspeech/SakSB14,SakIS2014,Sainath2015,peddinti2018low,povey2018semi}, whose primary advantage comes from their ability to model long temporal context. Output units for hybrid ASR are typically tied context-dependent (CD) states/phones, i.e., \textit{senones}, which are automatically clustered using decision trees \cite{Young:1994:TST:1075812.1075885} and require expertly-produced phonetic lexicons. There have been multiple attempts to model graphemes directly within the hybrid ASR framework, motivated by their simplicity \cite{killer2003grapheme,harwath2014speech,gales2015unicode,hadian2017towards,wang2018phonetic}. These efforts have been largely successful on low-resource languages, especially those whose written form encodes rich phonetic information \cite{killer2003grapheme,gales2015unicode}. However, on English where the grapheme-phoneme correspondence is poor and phonetic lexicons are highly optimized, grapheme-based approaches have consistently underperformed compared to phonetic baselines in terms of Word Error Rate (WER) \cite{hadian2017towards,wang2018phonetic}.
	
	With the rise of end-to-end techniques starting with Connectionist Temporal Classification (CTC) \cite{Graves:2006:CTC:1143844.1143891,Graves13hybridspeech,Graves:2014:TES:3044805.3045089,DBLP:conf/interspeech/SakSRB15,Amodei:2016:DSE:3045390.3045410}, followed by sequence-to-sequence attention-based models \cite{Chorowski:2015:AMS:2969239.2969304,chan2015listen,Prabhavalkar17,Chiu18,zeyer2018improved}, and sequence transducers \cite{Graves12transduction,Prabhavalkar17,Battenberg17RNNT,He2019RNNT}, the use of graphemic units has become more prevalent. These systems are able to directly model graphemes, word pieces, or whole words while achieving state-of-the-art performance on various ASR tasks. The ability to not rely on expert phonetic knowledge significantly reduces the development cost of new ASR systems and is often cited as a major advantage of end-to-end methods compared to traditional hybrid/HMM-based approaches. On the other hand, one disadvantage of end-to-end techniques is that they typically require larger amount of data to achieve good performance compared to hybrid methods. It is therefore appealing to combine the efficiency of hybrid ASR with the simplicity of graphemic modeling.
	
	In this paper, we reassess the assumption that hybrid ASR cannot model graphemes effectively, specifically for English. We show that contrary to popular beliefs, hybrid ASR systems utilizing tied CD graphemes, or \textit{chenones} for short, significantly outperform equivalent senone baselines by \textbf{4.5\%} to \textbf{11.1\%} relative on three different English datasets, the publicly available Librispeech corpus \cite{panayotov2015librispeech} and two large-scale in-house datasets in the Video and Assistant domains. Our chenone-based system achieves one of the best reported numbers to date on Librispeech, with WER of \textbf{3.2\%} on \texttt{test-clean} and \textbf{7.6\%} on \texttt{test-other} using only the provided 4-gram language model (LM) in decoding. We show that chenones can better exploit the increase in model capacity and training data compared to senones, leading to better recognition accuracy. Chenone-based systems also perform better at proper noun and rare word recognition, an area where senones are traditionally thought to have an advantage due to the use of human-curated lexicons and grapheme-to-phoneme (g2p) models. Finally, our ablative analysis shows that the key to achieving good performance with chenones is context- and position-dependent modeling. Based on these results, we conclude that the ability to model graphemes directly is not unique to end-to-end methods, and that traditional hybrid ASR systems can achieve better results by dropping all reliance on phonetic information.
	
	\section{RELATED WORK}
	\label{sec:related_work}
	
	Hybrid ASR systems have traditionally been built upon phonetic lexicons which map words to sequences of phones that encode their pronunciations. Phonetic lexicons, such as the CMU dictionary\footnote{\url{http://www.speech.cs.cmu.edu/cgi-bin/cmudict}}, are typically produced by linguists and undergo many careful reviews. We can further train a g2p model on these lexicons to predict pronunciations for previously unseen words \cite{bisani2008joint}. The main disadvantage of phonetic-based approaches is that such lexicons are difficult to create and maintain since they require specialized linguistic knowledge about the target language. The simplicity of graphemic modeling is therefore an appealing alternative.
	
	Previous work has shown that for several languages with a regular grapheme-phoneme relationship or complex segmental writing systems, graphemic modeling can perform on-par with or outperform phoneme-based approaches \cite{killer2003grapheme,gales2015unicode}. In \cite{gales2015unicode}, the authors proposed to derive phonetic features from the grapheme representation by extracting Unicode character descriptors; this enabled graphemic lexicons with CD modeling to outperform phonetic lexicons. By contrast, for languages that have a simple writing system with no explicit phonetic descriptor and an irregular grapheme-phoneme relationship, such as English, graphemic units have underperformed within the traditional HMM-based framework \cite{sung2009revisiting,kanthak2012,harwath2014speech,hadian2017towards,wang2018phonetic}. In \cite{hadian2017towards}, the authors explored end-to-end lattice-free MMI (LF-MMI) training of acoustic models. They showed that context-independent (CI) graphemes performed worse than CI phones on Wall Street Journal and Switchboard. A similar observation was drawn for both CI and CD modeling when benchmarking graphemes against phonemes on a multi-genre broadcast transcription task \cite{wang2018phonetic}. In \cite{sung2009revisiting}, the authors were able to get almost on-par performance using CD graphemes with letter-specific, coda, and onset modeling; however, this was done in the context of HMM-GMM trained with Maximum Likelihood instead of the HMM-DNN framework.
	
	The recent emergence of end-to-end techniques has enabled ASR systems to model graphemes directly while achieving state-of-the-art results on multiple English datasets \cite{DBLP:conf/interspeech/SakSRB15,Chiu18,park2019specaugment}. Within this paradigm, the modeling can be done at the grapheme level (e.g., ``\texttt{t h e \textunderscore \space c a t}") \cite{Graves:2006:CTC:1143844.1143891,chan2015listen,sainath2017no}, word piece level (e.g., ``\texttt{\textunderscore the \textunderscore c at}") \cite{zeyer2018improved,Chiu18,He2019RNNT}, whole word level (e.g., ``\texttt{the \textunderscore \space cat}") \cite{DBLP:conf/interspeech/SakSRB15,Soltau17,Audhkhasi18Word}, or a mixture of words and graphemes \cite{Li2018WordCTC,Ueno2018WordCTC}. Specifically for English, it has been shown that CI graphemes performed better than CI phones in sequence-to-sequence attention-based models, though the former under-performed on proper nouns and rare words \cite{sainath2017no}. Being able to remove the reliance on phonetic lexicons greatly simplifies the process of building new ASR systems and is often cited as a major advantage of end-to-end methods over the traditional hybrid approach.
	
	The contributions of this work are two-fold. Firstly, we present our approach to graphemic hybrid ASR which, for the first time, is able to consistently outperform equivalent phonetic baselines on a variety of English datasets. Our approach is based on well-known techniques in hybrid ASR with several modifications for graphemic units. This approach is an alternative for end-to-end ASR, combining the efficiency of traditional hybrid methods and the simplicity of grapheme-based modeling. Secondly, we provide detailed analysis to better understand the differences between phonetic and graphemic systems, including proper noun and rare word recognition accuracy, performance as a function of acoustic model (AM) capacity and amount of training data, as well as the importance of context and position dependency. Such in-depth studies have not been done on hybrid systems in previous work.
	
	\section{DATA}
	\label{sec:data}
	
	\subsection{Librispeech}
	\label{ssec:librispeech}
	
	Librispeech is a publicly available dataset consisting of audio book recordings (reading-style speech)\cite{panayotov2015librispeech}. The dataset contains 960 hours of acoustic training data, two development subsets (\texttt{dev-clean}, \texttt{dev-other}), and two test sets (\texttt{test-clean}, \texttt{test-other}). The ``other" type data is more acoustically challenging than ``clean" type data. All four development/test sets are between 5 to 5.5 hours in total duration. The official LM is a 4-gram model with 200K vocabulary trained on audio books (with much more text data than acoustic training transcripts). The official lexicon is a 200K phonetic lexicon with the same vocabulary as the LM, containing both human-curated pronunciations from the CMU dictionary\footnotemark[\value{footnote}] and g2p-generated pronunciations.
	
	\subsection{Video}
	\label{ssec:video}
	
	This dataset is sampled from in-house English video data publicly shared by users. The data are completely anonymized; both transcribers and researchers do not have access to any user-identifiable information (UII). The training set consists of 941.6K videos or 13.7K hours in length. We use two test sets, \texttt{vid-clean} with 1.4K videos (20.9 hours) and \texttt{vid-noisy} with 1.3K more acoustically and phonetically challenging videos (20.1 hours). All hyperparameter tuning is done on a held-out development set with 633 videos, totaling 9.7 hours in length. The average video is 52.6 seconds long with a standard deviation of 45.9 seconds. This dataset contains a diverse array of speakers, content types, and acoustic conditions, and is more challenging than the other two datasets considered in this work.
	
	\subsection{Assistant}
	\label{ssec:assistant}
	
	This in-house anonymized dataset is collected through crowd-sourcing. It consists of utterances recorded via mobile devices where crowd-sourced workers ask a smart assistant to carry out certain actions, such as calling a friend, playing music, or getting weather information. The training set comprises 15.7M utterances (12.5K hours) from 20K speakers. The development set for hyper-parameter tuning consists of 48 speakers disjoint from the training set, totaling 34.4K utterances (32.6 hours). The test set (\texttt{ast-test}) contains 58.8K utterances (50.4 hours) from 300 speakers that are unseen in both training and development. The average utterance length is 2.9 seconds with a standard deviation of 1.3 seconds.
	
	\section{GRAPHEMIC LEXICON FOR HYBRID ASR}
	\label{sec:graphemic_asr}
	
	\begin{table}[t]
		\centering
		\begin{tabular}{ l l }
			hello & \texttt{h\_WB e l l o\_WB} \\
			Michael's & \texttt{M\_WB i c h a e l ' s\_WB} \\
			Ritz-Carlton & \texttt{R\_WB i t z - C a r l t o n\_WB} \\
			DNN & \texttt{D\_WB N N\_WB} \\
			D.N.N. & \texttt{D\_WB N N\_WB} \\
			na{\"i}ve & \texttt{n\_WB a i v e\_WB} \\
		\end{tabular}
		\caption{Example entries in our proposed graphemic lexicon.}
		\label{table:graphemic_lexicon}
	\end{table}
	
	The primary challenge of graphemic ASR for English is that the mapping from graphemes to sounds is inherently ambiguous. Therefore, the key is to break down the output space with enough granularity for the AM to generalize effectively. Our proposed method is based on three main hypotheses. Firstly, we hypothesize that context dependency is especially important for graphemes. It is well-known in the ASR literature that senones significantly outperform CI phones due to the former's ability to account for co-articulation \cite{Young:1994:TST:1075812.1075885,Dahl2011}. Given that there is a high degree of ambiguity between graphemes and their acoustic realization, we argue that chenones will outperform CI graphemes by an even larger margin. This hypothesis is supported by previous findings, where CI phones outperform CI graphemes in the hybrid paradigm \cite{hadian2017towards,wang2018phonetic} and the improvement from CI to CD is larger for graphemes compared to phonemes \cite{wang2018phonetic}. Secondly, we hypothesize that position dependency is important for graphemes. This means that the same grapheme (e.g., ``a") may sound differently depending on whether it appears at the word boundary (e.g., ``amber", ``theta") or in the middle of a word (e.g., ``dart"). This is supported by previous experiments with HMM-GMM \cite{sung2009revisiting}; however, it is unclear if the result still holds in a hybrid setup. Thirdly, we hypothesize that casing information is important for graphemes. The convention in written English is to upper case abbreviations (e.g., ``DNN") and capitalize proper nouns (e.g., ``Michael"). We argue that when the data follow this convention, it is preferable to preserve the casing rather than lower-casing every letter. This may enable the model to better distinguish abbreviations from their lower-cased forms (e.g., ``SAT" vs. ``sat") and learn pronunciation variants of proper nouns. Combining context dependency, position dependency, and casing information will create enough granularity for the AM to handle the phonetic ambiguity of graphemes.
	
	Table \ref{table:graphemic_lexicon} gives several examples of what our graphemic lexicon looks like after incorporating these three hypotheses. The grapheme set used in this work is limited to the 26 standard English characters (both lower-cased and upper-cased), plus hyphens, apostrophes, and two special tokens, \texttt{SIL} and \texttt{GARBAGE}, which map to silence and out-of-vocabulary (OOV) words, respectively. Graphemes at word boundary positions are annotated with a special \texttt{WB} tag. The WB and non-WB versions of the same graphemes are technically separate acoustic units; however, they may be merged together during decision tree clustering. Letters that are not in the grapheme set are simply ignored; for example, ``DNN" and ``D.N.N." map to the same pronunciation since the grapheme ``." is skipped. We convert words with accent marks in them (e.g., ``na{\"i}ve") to their closest non-accented form using \texttt{unidecode}\footnote{\url{https://pypi.org/project/Unidecode/}}. Once the graphemic lexicon is prepared, we can apply traditional hybrid ASR techniques as usual, treating graphemes analogously as phonemes.
	
	\section{EXPERIMENTAL SETUP}
	\label{sec:experiments}
	
	\subsection{Lexicon Preparation}
	\label{ssec:lexicon}
	
	For graphemic lexicons, we follow the process described in Section \ref{sec:graphemic_asr}. The final grapheme set differs for each dataset due to different conventions in the training transcripts. Among the three datasets, Video is the only one where the training text has casing information; we use all lower-cased graphemes for the other two datasets.
	
	For phonetic lexicons, we follow the same approach to annotate phones at word boundaries with the \texttt{WB} tag. The Librispeech phonetic lexicon uses the provided CMU phone sets, preserving all stress markers. These different variants of the same phone may be clustered together during decision tree building. The phonetic lexicons for Video and Assistant use our in-house English phone set based on International Phonetic Alphabet (IPA), with no explicit stress markers.
	
	\subsection{GMM Bootstrapping and Decision Tree Building}
	\label{ssec:decision_tree}
	
	We train a bootstrap HMM-GMM system until Speaker Adapted Training (SAT), following the standard Kaldi Librispeech recipe\footnote{\url{https://github.com/kaldi-asr/kaldi/blob/master/egs/librispeech/s5/} (July 2019)}. This bootstrapping process uses 1K hours of randomly selected training data; for Librispeech this corresponds to the entire training set. We then generate alignments on the bootstrap data and build tri-phonetic/tri-graphemic decision trees with questions automatically generated from the alignment statistics. Each phoneme/grapheme and its word boundary (\texttt{WB}) variant can optionally share the same starting root node and may be clustered together. The number of tied CD phones (\textit{senones}) and graphemes (\textit{chenones}) ranges from 7K to 9K across all systems; this was chosen based on bootstrap WER on the development split. We model each phoneme/grapheme using a simple 1-state HMM topology with fixed self-loop and forward probability (0.5).
	
	\subsection{Acoustic Model Training}
	\label{ssec:acoustic_model}
	
	All experiments in this work employ multi-layer Latency-Controlled Bidirectional Long Short-Term Memory RNNs (LC-BLSTM) \cite{Zhang16LCBLSTM} with a softmax output layer trained on 80-dimensional log Mel-filterbank extracted with 25ms FFT windows and 10ms frame shift. The LC-BLSTM operates on 1.28s chunks with 200ms lookahead. The default model architecture has four hidden layers with 600 units for each direction (\texttt{4x600}), totaling approximately 35M free parameters. We also try two larger architectures for Librispeech, \texttt{5x800} (approx. 80M params), and \texttt{6x1000} (approx. 140M params), in order to better understand the AM performance as a function of network size. The first hidden layer of the LC-BLSTM subsamples the input by a factor of two \cite{peddinti2018low}, so that the posterior is emitted at a reduced 20ms frame rate.
	
	We employ speed perturbation \cite{Ko2015AudioAF} and SpecAugment's LD policy \cite{park2019specaugment} when training on Librispeech. Although these are considered data augmentation techniques, they do not use any additional resources other than the provided Librispeech data. We do not use any data augmentation for the other two datasets as we found no significant benefit from these techniques when the training set is large (more than 10K hours) and the model is small (less than 50M params).
	
	The AM training process happens in two stages. First we train the model with Cross Entropy (CE) loss for 25 epochs (Librispeech) or 20 epochs (Video and Assistant) with a batch size of 128, Adam optimizer \cite{kingma2014adam}, $5\times10^{-4}$ learning rate, 0.5 dropout, and Block-wise Model-Update Filtering (BMUF) with 0.875 block momentum \cite{Chen16BMUF}. The learning rate is halved whenever the development frame error rate (FER) does not improve. We use 16 GPUs during CE training for Librispeech and 32 GPUs for Video and Assistant. The best CE model in terms of development WER is used as the initial seed for LF-MMI \cite{povey2016purely} training in the second stage; we found that bootstrapping from CE gives slightly better results than training LF-MMI from scratch. For LF-MMI we train for 8 epochs with a batch size of 32, Adam optimizer, $10^{-5}$ learning rate, 0.5 dropout, BMUF with 0.875 block momentum, 0.1 CE interpolation weight, and a similar learning rate schedule. We use 24 GPUs during LF-MMI training for Librispeech and 48 GPUs for Video and Assistant. Unlike the original LF-MMI where training chunks are independent \cite{povey2016purely}, our training chunks have no overlap (except for the lookahead) and the forward LSTM states are carried over between contiguous chunks. The best LF-MMI model in terms of development WER will be used for final evaluation.
	
	\begin{table}[t]
		\centering
		\begin{tabular}{ | c | c | c | C C | }
			\hline
			\multicolumn{2}{|c|}{\textbf{Dataset}} & \textbf{Model} & \textbf{Ph} & \textbf{Gr} \\
			\hline
			\hline
			\multirow{6}{*}{Librispeech} & \textit{test-clean} & \multirow{2}{*}{\texttt{4x600}} & 4.0 & 3.8 \\
			& \textit{test-other} & & 9.3 & 9.6 \\
			\cline{2-5}
			& \textit{test-clean} & \multirow{2}{*}{\texttt{5x800}} & 3.8 & 3.4 \\
			& \textit{test-other} & & 9.0 & 8.4 \\
			\cline{2-5}
			& \textit{test-clean} & \multirow{2}{*}{\texttt{6x1000}} & 3.6 & \textbf{3.2} \\
			& \textit{test-other} & & 8.3 & \textbf{7.6} \\
			\hline
			\hline
			\multirow{2}{*}{Video} & \textit{vid-clean} & \multirow{2}{*}{\texttt{4x600}} & 16.1 & \textbf{15.0} \\
			& \textit{vid-noisy} & & 22.9 & \textbf{21.9} \\
			\hline
			\hline
			Assistant & \textit{ast-test} & \texttt{4x600} & 5.2 & \textbf{4.7} \\
			\hline
		\end{tabular}
		\caption{Word Error Rate (WER) comparison between phonetic (Ph) and graphemic (Gr) hybrid ASR systems.}
		\label{table:word_error_rate}
	\end{table}
	
	\begin{table}[t]
		\centering
		\begin{tabular}{ | c | c | c | F F | }
			\hline
			\textbf{System} & \textbf{AM} & \textbf{LM} & \textbf{test-clean} & \textbf{test-other} \\
			\hline
			\hline
			RWTH (hybrid) \cite{Luscher19RWTH} & 180M & 4g & 3.8 & 8.8 \\
			Kaldi TDNN-F \footnotemark[\value{footnote}] & 23M & 4g & 3.8 & 8.8 \\
			CAPIO (single) \cite{Han19Capio} & N/A & 4g & 3.6 & 8.9 \\
			\hline
			\hline
			Wav2Letter \cite{Liptchinsky17W2L} & 208M & 4g & 4.8 & 14.5 \\
			LAS + BPE \cite{zeyer2018improved} & N/A & 4g & 4.8 & 15.3 \\
			TDS Conv \cite{Hannun19TDS} & 37M & 4g & 4.2 & 11.9 \\
			NVIDIA Jasper \cite{li2019jasper} & 333M & 6g & 3.3 & 9.6 \\
			LAS + SpecAug \cite{park2019specaugment} & 360M & - & 2.8 & 6.8 \\
			LAS + SpecAug \cite{park2019specaugment} & 360M & 4g & 2.5 & 5.8 \\
			\hline
			\hline
			\textbf{Ours (Chenone)} & \textbf{140M} & \textbf{4g} & \textbf{3.2} & \textbf{7.6} \\ 
			\hline
		\end{tabular}
		\caption{Librispeech WER compared to published results, limited to single systems (no ensemble) without neural LM.}
		\label{table:librispeech_wer}
	\end{table}
	
	\begin{table*}[t]
		\centering
		\begin{tabular}{ c || D D }
			& \multicolumn{1}{c}{\textbf{Phonetic ASR Output}} & \multicolumn{1}{c}{\textbf{Graphemic ASR Output}} \\
			\hline
			\hline
			\multirow{5}{*}{\textbf{\textit{G}}} & then said sir \textbf{\color{red}fernando} there is nothing for it... & then said sir ferdinando there is nothing for it... \\
			& ...without disturbing the household of \textbf{\color{red}gain will} & ...without disturbing the household of gamewell \\
			& ...mademoiselle \textbf{\color{red}determination on} thinks... & ...mademoiselle de tonnay charente thinks... \\
			& ...and save us from the \textbf{\color{red}august} might & ...and save us from the ogre's might \\
			& ...in my morning room \textbf{\color{red}a jostling strock} & ...in my morning room at joscelyn's rock \\
			\hline
			\hline
			\multirow{5}{*}{\textbf{\textit{P}}} & ...the pre socratic philosophy are included... & the \textbf{\color{red}priests so critic} philosophy are included... \\
			& a great saint saint francis xavier & a great saint saint francis \textbf{\color{red}savior} \\
			& marmalades and jams differ little from... & \textbf{\color{red}margaret} and \textbf{\color{red}james} differ little from... \\
			& ...would then be given up to arsinoe & ...would then be given up to \textbf{\color{red}our sueno} \\
			& ...not my kind of haughtiness papa... & ...not my kind of \textbf{\color{red}fortune as} papa... \\
			\hline
		\end{tabular}
		\\
		\vspace{1em}
		{\textbf{\textit{G}}: graphemic performs better}\space\space\space$|$\space\space\space{\textbf{\textit{P}}: phonetic performs better}
		\caption{Example graphemic and phonetic ASR output on Librispeech test sets. Errors are indicated in \textbf{\color{red}red}.}
		\label{table:hypo_comparison}
	\end{table*}
	
	\subsection{Language Model and Decoding}
	\label{ssec:decoder}
	
	We use our in-house one-pass dynamic decoder with n-gram LM for all evaluations. The decoding parameters are tuned to minimize WER on the development set. For Librispeech, we use the official unpruned 4-gram LM with 200K vocabulary and 144M n-grams. For Video, we train a pruned 5-gram LM with 450K vocabulary and 35M n-grams on the training transcripts. For Assistant, we train a pruned 4-gram LM with 85K vocabulary and 23M n-grams on the training transcripts plus additional text data to increase the coverage. The LMs for Librispeech and Assistant are all lower-cased, whereas the one for Video preserves the original casing information.
	
	\section{RESULTS AND DISCUSSION}
	\label{sec:results}
	
	\subsection{WER Comparison: Phonemes vs. Graphemes}
	\label{ssec:graphemic_phonetic}
	
	Table \ref{table:word_error_rate} summarizes the WER of our phonetic (\textit{senone}) and graphemic (\textit{chenone}) hybrid ASR systems on Librispeech, Video, and Assistant. As can be seen, graphemic systems consistently outperform their phonetic counterparts on all three datasets. The relative WER improvement is larger on cleaner test sets, 8.4\%--11.1\% on Librispeech's \texttt{test-clean} and \texttt{test-other}, 7.0\% on Video's \texttt{vid-clean}, and 8.3\% on Assistant's \texttt{ast-test}, compared to 4.5\% on Video's \texttt{vid-noisy}. As shown in Table \ref{table:librispeech_wer}, our WERs on Librispeech are state-of-the-art compared to other hybrid models and competitive with end-to-end approaches. We could possibly get further improvement by incorporating neural LM rescoring \cite{Luscher19RWTH} and speaker adaptation\footnotemark[\value{footnote}].
	
	It is interesting to note that as we increase the AM size from \texttt{4x600} (35M params) to \texttt{6x1000} (140M params) for Librispeech, the graphemic system improves significantly, reducing WER by 15.8\% relative on \texttt{test-clean} and 20.8\% on \texttt{test-other}. By contrast, the relative improvement is smaller for the phonetic system, 10.0\% on \texttt{test-clean} and 10.8\% on \texttt{test-other}. This suggests that graphemic units may provide a more fine-grained output space that more powerful acoustic models are able to exploit.
	
	In order to better understand the differences between the two systems, we analyze Librispeech utterances in both \texttt{test-clean} and \texttt{test-other} where the graphemic system performs better and vice versa. As shown in Table \ref{table:hypo_comparison}, the phonetic system tends to misrecognize proper nouns with relatively poor pronunciations such as ``de [\texttt{D AH}] tonnay [\texttt{T AH N EY}] charente [\texttt{SH AA R EH N T EY}]", whereas the graphemic system typically does better on similar words. On the other hand, the graphemic system tends to fail on words whose grapheme sequences do not correspond well to how they are pronounced, such as ``xavier" and ``arsinoe." The graphemic system also frequently makes homophone errors where two words are spelled differently but sound similar, such as ``parlor" vs. ``parlour" and ``murdoch" vs. ``murdock." Developing methods to help graphemic systems rectify such errors will be an interesting future research direction.
	
	\subsection{Proper Noun and Rare Word Recognition}
	\label{ssec:proper_noun}
	
	\begin{table}[t]
		\centering
		\begin{tabular}{ | c | c | c | C C | }
			\hline
			\textbf{Dataset} & \textbf{Words} & \textbf{Split} & \textbf{Ph} & \textbf{Gr} \\
			\hline
			\multirow{4}{*}{Librispeech} & \multirow{2}{*}{Proper} & \textit{test-clean} & 7.4 & \textbf{6.0} \\
			& & \textit{test-other} & 16.5 & \textbf{14.4} \\
			\cline{2-5}
			& \multirow{2}{*}{Rare} & \textit{test-clean} & 7.6 & \textbf{7.0} \\
			& & \textit{test-other} & 18.0 & \textbf{16.2} \\
			\hline
		\end{tabular}
		\caption{Character Error Rate on proper nouns and rare words of phonetic (Ph) and graphemic (Gr) hybrid ASR systems.}
		\label{table:entity_char_error_rate}
	\end{table}
	
	The goal of this section is to objectively quantify the recognition accuracy on proper nouns and rare words on Librispeech, as a follow-up to the observation in Table \ref{table:hypo_comparison}. We first use an in-house named entity tagger to extract proper nouns from each test set. The number of tagged entities are as follows: \texttt{test-clean} (2.1K), \texttt{test-other} (2.2K). Here are some examples of extracted proper nouns: \textit{Buckingham}, \textit{Missouri}, \textit{Saint Paul}, \textit{John Calhoun}, \textit{Voltaire}, \textit{Leavenworth}, \textit{Jean Valjean}. We then align the ASR hypothesis against the reference text, and collect segments that are aligned with the tagged entities. Finally, we compute Character Error Rate (CER) on these hypothesis--reference segment pairs, which represents the system's error rate on proper nouns. We repeat this procedure to quantify CER on rare long-tail words, defined as words in the bottom 80\% in terms of frequency in the training set. This results in 1.3K and 1.4K selected words in \texttt{test-clean} and \texttt{test-other}, respectively, or 2.6\% of all words in the test sets.
	
	As shown in Table \ref{table:entity_char_error_rate}, graphemic systems clearly outperform phonetic baselines on proper noun and rare word CER on both test sets. This is rather surprising given that proper nouns and rare words were shown to be a weakness of end-to-end graphemic LAS models \cite{sainath2017no}. It could be that chenones, due to context and position dependency, are more conducive to accurate proper noun and rare word recognition than the CI graphemes used in their work. It will be an interesting follow-up study to see if end-to-end models can leverage chenones to improve their results further. It is also particularly interesting to compare chenones and word pieces \cite{zeyer2018improved,Chiu18,He2019RNNT} more closely. Both methods can be considered context-dependent modeling, but chenones leverage acoustic information while word pieces only utilize text data.
	
	\subsection{Ablative Analysis}
	\label{ssec:training_size}
	
	\begin{table}[t]
		\centering
		\begin{tabular}{ | c | c | c | C C | }
			\hline
			\textbf{Model} & \textbf{Hours} & \textbf{Dataset} & \textbf{Ph} & \textbf{Gr} \\
			\hline
			\multirow{8}{*}{\texttt{5x800}} & \multirow{2}{*}{50} & \textit{test-clean} & 7.4 & \textbf{7.3} \\
			& & \textit{test-other} & 18.0 & \textbf{17.8} \\
			\cline{2-5}
			& \multirow{2}{*}{200} & \textit{test-clean} & 5.0 & \textbf{4.6} \\
			& & \textit{test-other} & 11.9 & \textbf{11.8} \\
			\cline{2-5}
			& \multirow{2}{*}{480} & \textit{test-clean} & 4.3 & \textbf{3.8} \\
			& & \textit{test-other} & 9.9 & \textbf{9.5} \\
			\cline{2-5}
			& \multirow{2}{*}{960} & \textit{test-clean} & 3.8 & \textbf{3.4} \\
			& & \textit{test-other} & 9.0 & \textbf{8.4} \\
			\hline
		\end{tabular}
		\caption{Librispeech Word Error Rate of phonetic (Ph) and graphemic (Gr) ASR as a function of training data size.}
		\label{table:training_data_size}
	\end{table}
	
	
	\begin{table}[t]
		\centering
		\begin{tabular}{ | c | c c | c | C C | }
			\hline
			\textbf{Model} & \textbf{CD} & \textbf{PD} & \textbf{Dataset} & \textbf{Ph} & \textbf{Gr} \\
			\hline
			\multirow{6}{*}{\texttt{5x800}} & \multirow{2}{*}{N} & \multirow{2}{*}{Y} & \textit{test-clean} & \textbf{3.9} & 4.1 \\
			& & & \textit{test-other} & \textbf{9.4} & 10.7 \\
			\cline{2-6}
			& \multirow{2}{*}{Y} & \multirow{2}{*}{N} & \textit{test-clean} & 4.0 & \textbf{3.9} \\
			& & & \textit{test-other} & \textbf{8.9} & 9.2 \\
			\cline{2-6}
			& \multirow{2}{*}{Y} & \multirow{2}{*}{Y} & \textit{test-clean} & 3.8 & \textbf{3.4} \\
			& & & \textit{test-other} & 9.0 & \textbf{8.4} \\
			\hline
		\end{tabular}
		\caption{Librispeech Word Error Rate of phonetic (Ph) and graphemic (Gr) ASR under different context dependency (CD) and position dependency (PD) settings.}
		\label{table:cd_pd}
	\end{table}
	
	In this section, we analyze the performance of our ASR systems under different conditions to better understand the impact of various modeling decisions.
	
	We first investigate ASR performance as a function of \textbf{training hours}, where we limit LC-BLSTM training to \{50, 200, 480, 960\} hours of Librispeech data sampled randomly from the overall training set. Table \ref{table:training_data_size} shows that the difference between phonetic and graphemic systems starts out very small at 50 hours and gradually becomes larger as the data size increases. This suggests that graphemic AMs are better able to generalize given large amount of data, echoing the previous observation regarding network size in Section \ref{ssec:graphemic_phonetic} as well as the finding that graphemic system improved faster with more data \cite{sung2009revisiting}. One caveat with this analysis is that GMM bootstrapping and decision tree building were still done on the full 960 hours. Since these two steps are arguably more difficult for graphemes than phonemes, we may see the difference become smaller or even reversed if GMM bootstrapping and tree building are also done on the trimmed training sets.
	
	In terms of \textbf{casing information}, as hypothesized in Section \ref{sec:graphemic_asr}, we observe slightly better performance on Video (the only dataset with casing information) with cased graphemes compared to lower-cased graphemes: 15.0 vs. 15.4 on \texttt{vid-clean} and 21.9 vs. 22.0 on \texttt{vid-noisy}. Most of the improvement came from correctly recognizing abbreviations due to cased grapheme modeling.
	
	Finally, Table \ref{table:cd_pd} shows that \textbf{context} and \textbf{position dependency} are critical for graphemes, but not for phonemes. CI phones outperform CI graphemes by 4.9\% on \texttt{test-clean} and 12.1\% on \texttt{test-other}. The trend is reversed when we add context dependency (specifically tri-context); while phonetic results did not improve much, graphemic results improved significantly. This confirms that CD modeling is especially important for graphemes and is inline with findings in previous work \cite{hadian2017towards,wang2018phonetic}. Similarly, the phonetic system achieves similar performance with and without position dependency (using the \texttt{WB} tag), but the non-WB graphemic system degrades by 14.7\% on \texttt{test-clean} and 9.5\% on \texttt{test-other}. This means that the trend observed in HMM-GMM \cite{sung2009revisiting} also holds true for hybrid systems.
	
	\section{CONCLUSION AND FUTURE WORK}
	\label{sec:conclusion}
	
	In this paper, we establish that hybrid ASR systems utilizing chenones significantly outperform equivalent senone baselines in both overall WER and proper noun/rare word recognition. Powerful acoustic models, large training data, and context/position dependency are crucial to obtain optimal results with chenones. Based on these results, we argue that English hybrid ASR systems can be improved by removing the reliance on phonetic information, which in turn greatly simplifies the development process of new ASR models. Future work will explore using chenones with end-to-end techniques, improving graphemic results on long-tail words with unconventional spellings, and other graphemic modeling approaches beyond chenones.
	
	\bibliographystyle{IEEEbib}
	\newpage
	\small
	\bibliography{refs}
	
\end{document}